\documentclass[12pt]{JHEP3}
 \usepackage{epsf}

\usepackage{amsmath,amsfonts}

\newcommand{\tr}{\hbox{ Tr}}
\newcommand{\CO}{{\cal O}}
\newcommand{\bra}{{\langle}}
\newcommand{\ket}{{\rangle}}

\author{ David Berenstein $^{1,2,\dagger}$ and Samuel E. V\'azquez$^{1,\ddagger}$\\
$^1$ Department of Physics, UCSB, Santa Barbara, CA 93106 \\
$^2$ Kavli Institute for theoretical Physics, Santa Barbara, CA 93106
\\
$^\dagger$ \email{dberens@physics.ucsb.edu} $^\ddagger$
\email{svazquez@physics.ucb.edu}}

 \newcommand{\myfig}[3]{\begin{figure}[ht]
\begin{center}
\leavevmode
\epsfxsize=#2cm
\epsfbox{#1}
\end{center}
\caption{#3}
\label{fig:#1}
\end{figure}}

\title{ Integrable open spin chains from giant gravitons}
\abstract{We prove that in the presence of a maximal giant graviton state in ${\cal N}=4 $ SYM, the states dual to open strings attached to the giant graviton give rise to an $PSU(2,2|4)$ open spin chain model with integrable boundary conditions in the $SO(6)$ sector of the spin chain to one loop order. }

\preprint{hep-th/0501078\\
NSF-KITP-05-01}
\begin{document}

\section{Introduction}

The AdS/CFT correspondence is a non-perturbative equivalence between two different theories. One of them is given by string theory on an asymptotically AdS geometry as a theory of quantum gravity, and the
second one is an ordinary field theory on the boundary of the $AdS$ geometry. Because it is a non-perturbative equivalence, one can understand certain non-perturbative objects in string theory geometrically on the $AdS$ geometry, and we are interested in their dual realization in the field theory.

The main issue for understanding the dynamics of these objects is that geometric supergravity reasoning is only valid at strong 't Hooft coupling, whereas perturbative calculations in the field theory are usually only valid at weak 't Hooft coupling \cite{Malda}. At the level of perturbative expansions it is a strong-weak coupling duality for the 't Hooft coupling of the field theory.  If we concentrate on $AdS_5 \times S^5$, and the dual ${\cal N}=4 $ SYM theory with group $U(N)$ on the boundary,  the supergravity radius
of compactification in string units is of the order $R\sim (g_{YM}^2 N)^{1/4}$.

In order to compare both calculations we generally need to study objects which are not renormalized, or which receive small corrections, for which it is possible to extrapolate between weak and strong coupling. These objects are either BPS objects or almost BPS. Since we are interested in the study of non-perturbative objects, we want to analyze non-perturbative effects in the string theory which are protected. Moreover, we want to study D-brane states, as these usually have a simple description on the string worldsheet theory. This is because they give rise to boundary conditions for strings, but they do not affect the worldsheet sigma model to leading order.
 Fortunately we know that these objects exist in the $AdS_5\times S^5$ geometry. They are giant gravitons \cite{Giants}.

 Recently these objects have received renewed attention, because it is known how to describe the
 full back-reaction of the $AdS$ geometry in the supergravity limit and with arbitrary configurations of D-branes which preserve the same half of the supersymmetries \cite{LLM}. Also, these half-BPS states have a simple description in the dual ${\cal N}=4$ SYM theory in terms of free fermions for a one matrix model \cite{CJR,Ber}. This results from considering only the $s$-wave of one of the complex scalars of the ${\cal N}=4$ SYM on $S^3$.
 In this paper we want to understand how a single D-brane produces boundary conditions for strings ending on them.

As is well known, the classical  string theory  sigma model on $AdS_5\times S^5$ is integrable, and this is expected to persist at the quantum level \cite{BPR,Arut,Berk}.
Also, in the dual field theory, integrability also appears in the study of  the dual states (or single trace operators if one uses the operator sate correspondence of the CFT)  for string states on
$AdS_5\times S^5$, an observation which was first noticed in a subsector by Minahan and Zarembo \cite{MZ}, and later extended to the full supersymmetry group by Beisert and Staudacher \cite{BS}. See also \cite{DNW}.

For us, we will study the one loop anomalous dimensions of operators associated to strings ending on maximal giant gravitons. Other giant gravitons require a much more subtle analysis of the perturbation expansion which will be studied elsewhere. The main technical point is that the combinatorics for these other states behaves very differently than for maximal giants: the length of the spin chain is not constant
and this makes the details hard to understand.

For this paper  we will ask and answer the following two questions for the simple case we can calculate:
\begin{enumerate}
\item What is the full planar one loop spin chain model with boundary conditions (associated to the $SO(6)$ sector of SYM) for open strings ending on a maximal giant graviton?
\item Is this boundary spin chain model integrable or not?
\end{enumerate}

For the second question it is natural to conjecture that the
answer will be yes. This is mainly because the corresponding
D-brane is static and very symmetric in global coordinates. It is
then easy to find semiclassical solutions of open string ending on
the maximal giant by taking known  solutions \cite{FT} and cutting
the string so that the ends are on the giant graviton itself. A
more complete list of references and review material can be found
in \cite{Spinref}. We will show integrabiltiy by studying
scattering of ``defects" with respect to a Bethe ground state from
the boundary, and showing that we can satisfy a boundary Yang
Baxter equation for these defects. Recently it has been  argued
that this is the easiest way to ascertain the integrability of
spin chain models associated to ${\cal N}=4$ SYM theory
\cite{Staud} beyond one loop. In our case, this is applied to the
case of a boundary spin chain model, and we are studying
dispersionless scattering from the boundary and it's compatibility
with the scattering between defects.

Other setups with integrable boundary conditions for strings ending on non-compact branes (defect D-branes or flavour branes)  have been analyzed in the literature already \cite{open}. These are characterized by having some type of quark (flavor matter) at the ends of the spin chain. In these cases the dual theory has more degrees of freedom on the boundary than just ${\cal N}=4 $ SYM. In this paper we are studying non-perturbative objects which are accessible as a state of the field theory which is not close to the perturbative vacuum. In this sense, we are exploring how the 't Hooft expansion rearranges itself in the presence of large perturbations away from the vacuum, whose energy is of order $N$.

Finally, these states have a similar combinatorial structure to baryons, and this field theory exercise has to be seen as a toy model to systematically study these objects as well.

\section{Setting up the perturbative problem}

Our purpose is to understand in detail the boundary conditions that a particular D-brane state
(the maximal giant graviton) imposes on the $SO(6)$ integrable spin chain derived from the
${\cal N}=4 $ SYM by Minahan and Zarembo, and further extended to the full $SU(2,2|4)$
by Beisert and Staudacher.

To do this, we need to understand precisely the dictionary between string states attached to the giant graviton and the dual CFT description. Part of this program has been spelled out already for maximal
giant gravitons in previous work \cite{maxgiants,tech} and we can be certain that the dictionary is understood because
one can show that the DBI quadratic fluctuations of the D-branes \cite{vib} can be reproduced in the large $N$ limit \cite{tech}. For other sphere giants this has not been carried out in detail yet. This is not accidental: smaller giant gravitons move, and they drag string states attached to them, so that some of the momentum that the string carries should be along the direction of motion of the giant graviton. If one wants to compare the field theory analysis with the supergravity calculation, this effect has to be taken into account properly.
Now we concentrate on the case for the maximal giant graviton.

The important point is that the operators dual to these giant gravitons are given by determinant like operators \cite{maxgiants,CJR}, and attaching a string to the giants gives an operator of
the form
 \begin{eqnarray}
    \label{giantop}
{\cal O}(x) = \epsilon^{j_1 \cdots j_N}_{i_1 \cdots i_N}
Z_{j_1}^{i_1} \cdots Z_{j_{N-1}}^{i_{N-1}} (\phi_1 \phi_2 \cdots
\phi_L)_{j_N}^{i_N}\;,
\end{eqnarray}
where $\phi_i$ can be one of the fields: $X, Y, Z, \overline{X},
 \overline{Y}$, or ${Z}$. In principle one can generalize the spin
chain to include covariant derivatives, spinors, etc. These in
some sense contain the information of $AdS_5$, where the giant
graviton looks like a point-like object moving in time.
 The six scalar directions contain the information of the string on $S^5$. This is also the situation where most of the semiclassical
string states have been understood and related to the perturbative calculations \cite{GKP,FT,Spinref,Kruc}.

In the above, the giant graviton is built mostly out of $Z$, and at the end we intertwine the indices of a word to the indices of the giant graviton operator.
It is this string of concatenated $\phi$ where the spin chain resides, and the length of the word is $L$.

Next, we need to calculate the planar anomalous dimensions of the spin chain and understand the
boundary conditions that are generated by interactions with the $Z$. Also notice that there is a
boundary
constraint that $\phi_1$ and $\phi_L$ have to be different than $Z$. Otherwise the operator factorizes into a closed string plus a maximal giant graviton \cite{maxgiants}.

Now that we have the set of operators that we are going to study, we need to calculate the one loop anomalous dimension matrix of this collection of operators with the above index structure and where
we consider all possible fixed values of the different strings of length $L$ in the planar diagram approximation.

Because of the presence of the D-brane state, we need to be careful in
 characterizing contributions as planar/non-planar when we also include contractions with the field
 $Z$ that come from the determinant part of the operator.

 This requires some combinatorial ability, but in the end it is straightforward once the combinatorics is
 taken care of. The basic combinatorics has already been discussed in the appendices of the following references \cite{tech} and we will use these available results in what follows.   An important observation is that one can systematically write the calculations in terms of a series of free matrix model correlators, as these just depict the group index contractions required to glue propagators together .  Here we just need to borrow some of the combinatorial structures found in this way and do the corresponding one loop calculations in detail.

We begin by defining the correlation function,
\begin{eqnarray}
M_{\alpha \beta} = \left \langle \tilde{\cal O}^*_\alpha(x)
\tilde{\cal O}_\beta(0) \right \rangle_{free + interacting} \;,
\end{eqnarray}
where,
\begin{eqnarray}
\tilde{\cal O}_\alpha(x) = \frac{{\cal O}_\alpha(x)}{\left \langle
{\cal
 O}^*_\alpha(x) {\cal O}_\alpha(0) \right \rangle_{free}^{1/2}}\;,
\end{eqnarray}
and where the indices $\alpha$, $\beta$ denote different open
string configurations.  At one-loop and in the large $N$ 't Hooft
limit, $M_{\alpha \beta}$ will have the following general form,
\begin{eqnarray}
\label{Mexp}
 M_{\alpha \beta} = \delta_{\alpha \beta} -
2 \, \Gamma_{\alpha \beta} \log(|x|\Lambda) + \ldots \;,
\end{eqnarray}
where $\Gamma_{\alpha \beta}$ is the matrix of anomalous dimension
and $\Lambda$ is an ultraviolet cutoff. We then identify the
anomalous dimension matrix with the Hamiltonian of the
corresponding string quantum states,
\begin{eqnarray}
\label{gamma}
 \Gamma_{\alpha \beta} \cong \langle \psi_\alpha
|H|\psi_\beta \rangle\;,
\end{eqnarray}
where $\tilde{\cal O}_\alpha \cong |\psi_\alpha \rangle$.
This is just the operator state correspondence which is available for any CFT: the Hamiltonian
corresponds to taking radial time and compactifying the CFT on a round sphere.

The relevant part of the ${\cal N} = 4$ SYM action for our calculation is just the bosonic sector, because we are looking only at the $SO(6)$ part of the spin chain:
\begin{eqnarray}
\label{action} S = \frac{1}{2 \pi g_s} \int d^4x
\textrm{Tr}\left(\frac{1}{2} F_{\mu \nu} F^{\mu \nu} + D_\mu X
D^\mu \overline{X} + D_\mu Y D^\mu \overline{Y} + D_\mu Z D^\mu
\overline{Z} + V_D + V_F\right)\;, \nonumber \\
\end{eqnarray}
where,
\begin{eqnarray}
V_D &=& \frac{1}{2} \textrm{Tr}\left( | [X, \overline{X}] + [Y,
\overline{Y}] + [Z, \overline{Z}] |^2 \right)\;, \\
V_F &=& 2 \textrm{Tr}\left( |[X,Y]|^2 + |[X,Z]|^2 + |[Y,Z]|^2
\right)\;.
\end{eqnarray}

The problem naturally splits into two parts. First, for the bulk of the spin chain, the one loop contributions are known and were derived by Minahan and Zarembo \cite{MZ}. For our case, we neeed to consider the one loop contribution for all six possible fields $\phi$ at the boundary. For $X,\bar X, Y, \bar Y$ the results are the same because one can relate these four states to each other using the residual $SO(4)$ symmetry that
the giant graviton preserves. However for $Z$ we are not allowed to put it on the first site and
for $\bar Z$, it interacts differently with $Z$ than the other four fields, so it needs to be treated separately.

\subsection{Working the combinatorics}

First, we need to do the free field contractions between different configurations, in order to calculate the norms of the states.

If we use  the composite matrix word $W^i_j= (\phi_1\dots \phi_L)^i_j$, and label the operators by the words  $W$ themselves, as $\CO_W$, we can do  the free field contractions by steps. First the $Z$ are paired with the $\bar Z$ in the determinants, and then we contract the rest. For simplicity we will use contractions $\bra Z^i_j,\bar Z^k_l\ket \sim \delta^i_l\delta^k_j$, and forget the spacetime dependence. This can always be reinserted back into the calculation at the end. In fact, all integrals that need to be performed are available in the literature \cite{BMN}, and they are of the form
\begin{equation}
\int d^4 x \frac{1}{|x-y|^4 |x-y'|^4}\sim \frac{1}{|y-y'|^4}
\log(|y-y'|)\;.
\end{equation}
 What is not available is the combinatorics of the Feynman  diagrams, which is what we we will concern ourselves with.

 The contribution we are interested in is just the combinatorics of contractions. Because we have forgotten the spacetime dependence, these can be obtained considering a free
 matrix model with one matrix for each of the six scalar fields \footnote{This can also be considered as a dimensional reduction on the  s-wave of the scalar fields on $S^3$, a point raised in \cite{Ber}}
\begin{equation}
\bra \bar \CO_{W'} \CO_W \ket \sim (N-1)!^3 \bra \tr(\bar {W'} W)
\ket \sim (N-1)!^3\delta_{W',W} N^{L+1} \; .
\end{equation}
The $(N-1)!^3$ comes from having $(N-1)!$ possible permutations of the $Z$ in the determinant, and from the identity
\begin{equation}
\epsilon^{i_1,\dots ,i_{N-1},i}\epsilon_{i_1, \dots i_{N-1}, j}(N-1)! \delta^i_j\;. \label{eq:oneindex}\end{equation} The second
part of the calculation that gives $N^L$ is just the planar
contribution to the norm of the state in the matrix model. All
other contributions to the normalization are subleading in $1/N$.

The one loop anomalous dimensions has two types of contributions. Those from the bulk of the spin chain, where all the $Z$ that appear as the bulk of the determinant
operator $\CO_W$ are not attached to the interaction vertex, and those where one of the $Z$ is attached to the interaction vertex.

It is easy to show that  the first ones just give the spin chain Hamiltonian in the bulk as computed by Minahan and Zarembo.

To do the one loop anomalous dimension calculation interacting with the boundary, we need to keep one of the $Z$ in the initial operator without doing the Wick contraction with the corresponding $\bar Z$.
These will be contracted with the interaction terms in the Lagrangian. This can be represented graphically as in figure \ref{fig:OneZun.eps}.

\myfig{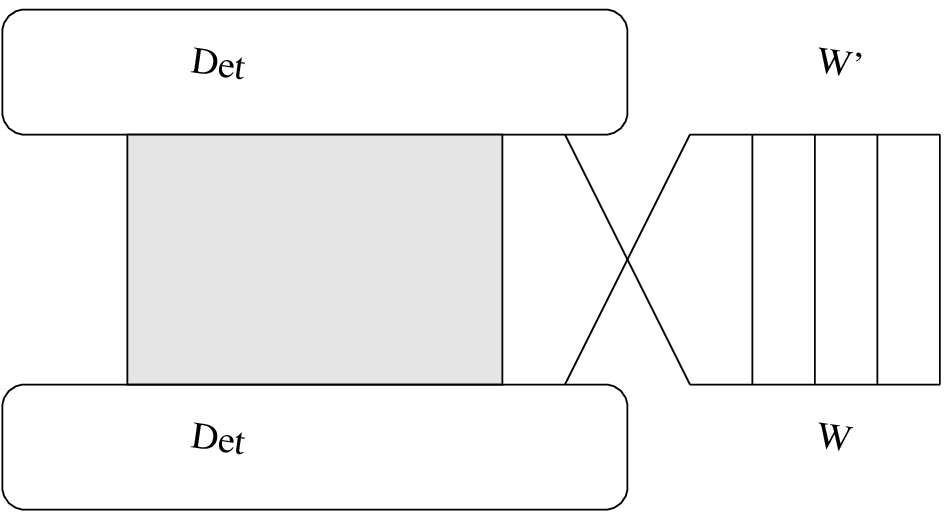}{6}{Pictorial description of contractions leading to boundary conditions of the spin chain: The round boxes represent the determinant part of the operator, and the straight horizontal lines
represent the spin chain part of the problem. Vertical lines indicate single contractions, and the thick vertical filled box represents the contractions of $Z,\bar Z$ between the determinant part of the operators. }

Doing the rest of the contractions we get a collection of different trace structures which can be divided in left and right boundary conditions for the string
\begin{equation}
\bra \bar \CO_{W'} \CO_W \ket _{\hbox{One $Z, \bar Z$
uncontracted}} \sim (N-2)!^3 (N-1)^2 \bra
\delta^{ij}_{kl}\delta^{i'j'}_{k'l'} Z_i^{k} W_j^l \bar
Z_{i'}^{k'} \bar {W'}_{j'}^{l'}   \ket \;.
\end{equation}
The $(N-1)^2$ results from choosing one $Z$ and one $\bar Z$ from
the operators $\CO_W$ and $\bar \CO_{W'}$ that will be used in the
interaction. The
$\delta^{ij}_{kl}=\delta^i_k\delta^j_l-\delta^i_l\delta^j_k$ is
totally antisymmetric in the upper and lower indices and results
from identities like (\ref{eq:oneindex}) with more indices
uncontracted, and it is easy to see that we are getting an
expansion in traces of different combinations of the $Z, W, \bar
Z, \bar W$. Depending on the order of the fields, the $\bar Z$
from the determinant will appear either to the left or to the
right of the word $W$, and from the ordering we can observe the
boundary conditions.
 In the end, there are only two important orderings for the leading planar diagrams:
$\tr(W\bar Z Z\bar W') $ and $\tr(W \bar W' Z\bar Z)$, and one is interpreted as the left boundary condition, while the other one is interpreted as a right boundary condition.

We have to distinguish two cases. Where the first letter of $W$ is one of $Y,\bar Y, X, \bar X$ and where it is a $\bar Z$. In the end, we also need to compare the ratio of the normalization of the boundary term we get with the
ones on the bulk of the spin chain.
For $Y$ or $X$ as the first letter, we can exploit the chiral nature of the operator, and it is known that in
this case only $F$ terms contribute: the D-term and photon exchange cancel against each other. Incidentally, this can be used to determine the photon exchange diagram, which is flavor blind, but more complicated as we need two vertices.

The F terms end up having an ordering where two chiral operators
are followed by two anti-chiral ones. This is from $V_F=\tr
[(YZ-ZY)(\bar Y\bar Z-\bar Z\bar Y)]$. Looking at the contractions
in the operator $Z$ and $Y$ have to be separated either by  $\bar
Z$ or $\bar W'$, so the leading planar $F$ term correction
vanishes. This is just as it should be. If the word $W$ is just
$Y$, then the operator is a descendant of $\det(Z)$, and the
anomalous dimension vanishes. This means that the boundary
contribution for $Y$ from the left and right has to be zero. As
argued above, by the $SO(4)$ symmetry, this is true also for $\bar
Y$ and $\bar X$. Therefore the boundary Hamiltonian for any of
these fields vanishes.

Now we can also put the $\bar Z$ on the boundary. Since the operator is not chiral,
 the photon exchange and $D$ term don't cancel. Instead, they add up.
 Also, one can see from the $D$ terms that the combination that appears has terms with the same ordering as the terms we get in the calculation. Take the word $W= \bar Z W_0$, then we get
 \begin{equation}
 \frac 14 \bra \tr( \bar  W_0' Z Z \bar Z \bar Z W_0) \tr([Z,\bar Z]^2)
 \ket\;.
 \end{equation}
 The contribution with the ordering $\bar Z^2 Z^2$ in the interaction gives us a factor of $1/2$.
 We get another $1/2$ from the photon exchange, to get a planar anomalous dimension which is normalized to $1$\footnote{The $D$ term generically contributes with a factor of $1/2$ with respect to the F-terms because it is real. Also, the D-term for the $Z,\bar Z$ contributes with the opposite sign than the $D$ term for the chiral case, because we need to take different orderings of the expansion of the commutators}. This is in the same normalization that the bulk spin chain Hamiltonian nearest neighbor exchange term has coefficient $(-1)$. This is
 \begin{equation}
 H_{l,l+1} \sim \frac12[ K_{l,l+1} +
2 (I_{l,l+1} - P_{l,l+1})] \;,
 \end{equation}
and in the above equation the exchange term is the permutation operator $P_{l,l+1}$.

This concludes our calculation of the spin chain hamiltonian with
it's boundary conditions: the boundary term for the fields $X,Y,
\bar X, \bar Y$ is zero, while for $\bar Z$ it is $1$ with
positive sign. As discussed before, we can not put a $Z$ at the
beginning or end of $W$, so in the Hamiltonian above we impose
this constraint on the possible states of the theory. Also, we can
now reinsert the 't Hooft coupling into the spin chain Hamiltonian
with the appropriate numerical constants. At the end we obtain,
\begin{eqnarray}
\label{refH} H &=& \frac{1}{2}\lambda \sum_{l = 2}^{L-2}
[K_{l,l+1} + 2
(I_{l,l+1} - P_{l,l+1})] \nonumber \\
&&+ \frac{1}{2}\lambda \,Q_1^Z[K_{1,2} + 2 (I_{1,2} -
P_{1,2})]Q_1^Z \nonumber \\
&&+ \frac{1}{2}\lambda \,Q_L^Z[K_{L-1,L} + 2
(I_{L-1,L} - P_{L-1,L})]Q_L^Z \nonumber \\
&&+ \lambda (I - Q_1^{\overline Z}) + \lambda (I - Q_L^{\overline
Z})\;,
\end{eqnarray}
where we have defined $\lambda \equiv \frac{g_s N}{2 \pi} =
\frac{g_{YM}^2 N}{8 \pi^2}$, and the operator $Q_l^\phi$ acts in
the following way: $Q_1^\phi|\phi\cdots\rangle = 0$,
$Q_1^\phi|\varphi\cdots\rangle = |\varphi\cdots\rangle$ for
$\varphi \neq \phi$, and similarly for $Q_L^\phi$. This is,
$Q_1^\phi$ is a projection on the orthogonal complement of the
state $\phi$ in the first site of the spin chain.

In the next section we will ask and determine wether the above spin chain Hamiltonian with boundary conditions is indeed integrable or not.

\section{Scattering at the boundary}

Now that we have our spin chain Hamiltonian, we want to ask if it
is integrable or not. In principle there are various ways to do
this. The easiest way by far is if one is able to use a Bethe
ansatz  solution of the problem, and this is the route we will
take \footnote{A useful set of references about integrable systems
and the Bethe ansatz is \cite{Bethe, Faddeev1, Faddeev2, intsys,
Doikou, book, qgroups}.}. We can do this because we have an
unbroken $SO(4)$ symmetry and we can look for a Bethe ground state
as the highest possible weight under this $SO(4)$ group. This
state is unique. The word we need as the ground state is $Y^L$.
Incidentally, this is also the route to the plane wave limit if we
also scale $L \sim \sqrt N$, and which has been analyzed elsewhere
\cite{maxgiants}.

The idea now is to introduce defects in the word $Y^L$. These can be either $Z,\bar Z, X, \bar X$.
In our case inserting a $\bar Y$ is considered as a bound state of $X,\bar X$ for example, because it will generically mix with states containing $X,\bar X$ next to each other. This means we do not need it when we are considering single defect states.

Let us analyze the Bethe ansatz for each of these single defects. Naturally, there is reflection at the boundary. Conservation of energy implies that the boundary only contributes a phase shift for each of these particles, plus the change in sign of the momentum. In principle different states could mix, but the symmetry of the boundary conditions prevent this, so we can analyze this scattering directly.
This gives us what we call the boundary S-matrix.

To calculate it, we consider states with a single impurity. We
start with a single $X$ or $\overline X$ excitation. The most
general energy eigenstate is,
\begin{eqnarray}
|\psi\rangle = \sum_{x=1}^{L}f(x)|x\rangle\;,
\end{eqnarray}
where,
\begin{eqnarray}
\label{fx} f(x) = A e^{i k x} + \tilde{A} e^{-i k x}\;,
\end{eqnarray}
and $|x\ket = |YY\cdots Y\phi Y\cdots Y\ket$ on the spin chain
with $\phi =   X, \overline X$ at position $x$.

For these impurities, the Hamiltonian acts as follows on the states
\begin{eqnarray}
H|x\rangle = \lambda (|x\rangle - |x-1\rangle)(1 - \delta_{x,1}) +
\lambda (|x\rangle - |x\rangle)(1 - \delta_{x,L})\;.
\end{eqnarray}
or in matrix form
\begin{equation}
H = \lambda \begin{pmatrix} 1& -1 & 0 &0& \dots\\
-1& 2& -1 & 0& \dots\\
0& -1 & 2& -1&\dots\\
\vdots& \ddots& \ddots &\ddots &\vdots
\end{pmatrix} \;.
\end{equation}

So, as before, the condition that $|\psi\rangle$ is an eigenstate
of $H$ translates to
\begin{eqnarray}
\label{xXenergyeq}
 E(k) f(x) = \lambda (f(x) - f(x-1))(1 -
\delta_{x,1}) + \lambda (f(x) - f(x+1)(1 - \delta_{x,L})\;.
\end{eqnarray}
For $x \neq 1, L$ Eq.(\ref{xXenergyeq}) gives the expected energy
of a single impurity,
\begin{eqnarray}
\label{singleE}
 E(k) = 2\lambda (1 - \cos k)\;.
\end{eqnarray}
For $x = 1$, Eq. (\ref{xXenergyeq}) gives
\begin{eqnarray}
f(1) = f(0)\;,
\end{eqnarray}
which implies,
\begin{eqnarray}
\label{bSmxX} \frac{A}{\tilde{A}} = - \frac{1 - e^{-i k} }{1 -
e^{i k}}\;.
\end{eqnarray}
For $x = L$, the second boundary,  we get
\begin{eqnarray}
\label{Lboundary}
 f(L+1) = f(L)\;.
\end{eqnarray}

Now we consider a $Z$ excitation. The Bethe ansatz for the energy
eigenstate is,
\begin{eqnarray}
|\psi\rangle = \sum_{x=2}^{L-1}g(x)|x\rangle\;,
\end{eqnarray}
where,
\begin{eqnarray}
\label{fz} g(x) = B e^{i k x} + \tilde{B} e^{-i k x}\;,
\end{eqnarray}
and $|x\rangle$ has the obvious definition. The Hamiltonian for
this impurity reduces to
\begin{eqnarray}
H = \lambda \sum_{l = 2}^{L-2} (I_{l,l+1} - P_{l,l+1}) + \lambda
Q_1^Z(I_{1,2} - P_{1,2})Q_1^Z + \lambda Q_L^Z(I_{L-1,L} -
P_{L-1,L})Q_L^Z\;. \nonumber \\
\end{eqnarray}
which in matrix form reads
\begin{equation}
H = \lambda \begin{pmatrix}
 2& -1 & 0 &0& \dots\\
-1& 2& -1 & 0& \dots\\
0& -1 & 2& -1&\dots\\
\vdots& \ddots& \ddots &\ddots &\vdots
\end{pmatrix}\;.
\end{equation}
In the above, the first possible case with
$Z$ at the boundary is projected out, so we only get the bulk spin chain Hamiltonian without
the hopping term to the first site.

Following the same procedure as above, we obtain an the following
equation,
\begin{eqnarray}
E(k) g(x) &=& \lambda (g(x) - g(x-1))(1 - \delta_{x,2}) \nonumber
\\
&&+ \lambda
(g(x) - g(x+1))(1 - \delta_{x,L-1}) \nonumber \\
&&+ \lambda\, \delta_{x,2}\, g(x) + \lambda\, \delta_{x,L-1}\,
g(x)\;.
\end{eqnarray}

For $x \neq 2, \, L-1$ we get the same energy as in
(\ref{singleE}). Using this for $x = 2$ we get
\begin{eqnarray}
g(1) = 0\;,
\end{eqnarray}
which implies,
\begin{eqnarray}
\label{bSmz} \frac{B}{\tilde B} = - e^{- 2 i k}\;.
\end{eqnarray}
For $x = L-1$ we get,
\begin{eqnarray}
g(L) = 0\;,
\end{eqnarray}
This spin wave sees an effective chain of length $L - 2$ and
vanish one step pass the ends of this chain. Hence, we see that
this state has Dirichlet boundary conditions.

Finally, we look at the case of a $\overline Z$ impurity. The
Hamiltonian is,
\begin{eqnarray}
H = \lambda \sum_{l = 1}^{L-1} (I_{l,l+1} - P_{l,l+1}) + \lambda
(I - Q_1^{\overline Z}) + \lambda (I - Q_L^{\overline Z})\;.
\end{eqnarray}
The eigenstate has the form,
\begin{eqnarray}
|\psi\rangle = \sum_{x=1}^{L}h(x)|x\rangle\;,
\end{eqnarray}
where,
\begin{eqnarray}
\label{hx} h(x) = C e^{i k x} + \tilde{C} e^{-i k x}\;.
\end{eqnarray}

 Following the usual procedure we
obtain,
\begin{eqnarray}
E(k) h(x) &=& \lambda (h(x) - h(x-1))(1 - \delta_{x,1}) \nonumber
\\
&&+ \lambda
(h(x) - h(x+1))(1 - \delta_{x,L}) \nonumber \\
&&+ \lambda\, \delta_{x,1}\, h(x) + \lambda\, \delta_{x,L}\,
h(x)\;.
\end{eqnarray}
and in matrix form we obtain the same result as for the case of $Z$, but $x$ is allowed to start at $x=1$ as opposed to $x=2$.

At $x \neq 1, L$ we get the usual dispersion relation. For $x = 1$
we have that the boundary condition reduces to
\begin{eqnarray}
h(0) = 0\;.
\end{eqnarray}
For $x = L$ we get,
\begin{eqnarray}
h(L+1) = 0\;,
\end{eqnarray}
So the wave function vanishes at $0$ and $L+1$, and we also get Dirichlet boundary conditions. The boundary condition at $x=0$ is easy to read, and we obtain
\begin{equation}
C/{\tilde C}=-1\label{bSmZ} \;.
\end{equation}

We now have calculated the scattering of all one particle defects from the boundary.
All of these can be encoded in a single matrix $K$ which is diagonal and describes how the right moving modes are related to the left moving modes,
\begin{equation}
|\psi (-k)\rangle = K(k) |\psi (k)\rangle\label{bSm} \;.
\end{equation}
where $K(k)$ is the relation $(A,B,C) (k) = K(k) (\tilde A,\tilde B, \tilde C) (k)$ above.

Armed with this, we need to examine a two particle Bethe ansatz. The wave function of the two particle Bethe ansatz has more components than in the case of a periodic boundary condition because we need to take into account the reflections of momenta from the boundary.
 Setting this up also requires understanding the $2\to 2$ scattering of impurities.

 If we ignore the boundary conditions, one can calculate a $2\to 2$ scattering matrix $S$ by
 looking at states of the schematic form
 \begin{equation}
 |k_1,k_2> = \sum_{x_1<x_2} \exp(ik_1 x_1+ik_2 x_2) |x_1,x_2>
 +\sum_{x_1<x_2} S \exp(ik_2 x_1+ik_1 x_2)|x_1,x_2> \;.
 \end{equation}
where in the above we have omitted the group theory quantum numbers for $SO(6)$ on the description of the states. Also, if we have a singlet wave function with respect to $SO(4)$
one also needs to include the ``one impurity" wave function associated to the state $\bar Y$
with which the state can mix. This is necessary to calculate the matrix $S$, but once $S$ is obtained, we can ignore it to some extent.

Let $|x_1,x_2\rangle_{\phi_1 \phi_2}$ be a state with two
impurities $\phi_1$ and $\phi_2$ located at positions $x_1$ and
$x_2$ respectively along the spin chain ($x_1 < x_2$). We can
divide these states in two sets: the ones where $\phi_1 \neq
\overline\phi_2$ and the ones with $\phi_1  = \overline\phi_2$. It
is easy to see that the Hamiltonian will never mix both sets.

Following the Bethe ansatz, we can write the most general
eigenstate as,
\begin{eqnarray}
|\psi\rangle = \sum_{ x_1 < x_2 } f_{\phi_1 \phi_2}(x_1,x_2)
|x_1,x_2\rangle_{\phi_1\phi_2} + \sum_{ x_1 < x_2 } f_{\phi_2
\phi_1}(x_1,x_2) |x_1,x_2\rangle_{\phi_2\phi_1}\;,
\end{eqnarray}
where
\begin{eqnarray}
 f_{\phi_1\phi_2}(x_1,x_2) &=& A_{\phi_1\phi_2}(12) e^{i(k_1 x_1 + k_2 x_2)} +
 A_{\phi_1\phi_2}(21)
e^{i(k_2 x_1 + k_1 x_2)} \;, \\
 f_{\phi_2\phi_1}(x_1,x_2) &=& A_{\phi_2\phi_1}(12) e^{i(k_1 x_1 + k_2 x_2)} +
 A_{\phi_2\phi_1}(21)
e^{i(k_2 x_1 + k_1 x_2)}\;.
\end{eqnarray}

For this first case, after some algebra,  we get
\begin{eqnarray}
\label{S1}
\left(%
\begin{array}{c}
  A_{\phi_1\phi_2}(12) \\
  A_{\phi_2\phi_1}(12) \\
\end{array}%
\right) = \left(%
\begin{array}{cc}
  u - 1 & u \\
  u & u - 1 \\
\end{array}%
\right)\left(%
\begin{array}{c}
  A_{\phi_1\phi_2}(21) \\
  A_{\phi_2\phi_1}(21) \\
\end{array}%
\right)\;,
\end{eqnarray}
where,
\begin{eqnarray}
u(k_2,k_1) = \frac{ e^{ik_1} -e^{ik_2}}{1 + e^{ik_2}(e^{ik_1 } -
2)}\;.
\end{eqnarray}

Now we consider the case of states of the form
$|x_1,x_2\rangle_{\phi \overline\phi}$. Of course, these states
will also include $|x_1\rangle_{\overline Y}$ which is a
$\overline Y$ located at position $x_1$ along the spin chain. A
state carrying such an impurity has the same R-charge as a state
with a pair of the form $\phi\overline\phi$. Therefore, a single
$\overline Y$ can be visualized as a bound state of the form
$\phi\overline\phi$. Henceforth we will denote an holomorphic
impurity ($X$ or $Z$) by a $\phi$ and a antiholomorphic impurity
($\overline X, \overline Y$ or $\overline Z$) by a $\overline
\phi$. The Hamiltonian in this case takes the full form
(\ref{refH}). The Bethe ansatz for the energy eigenstates is,
\begin{eqnarray}
|\psi\rangle = \sum_{ x_1 < x_2 } f_{\overline Y}(x_1)
|x_1\rangle_{\overline Y} + \sum_{\phi = X, Z}\sum_{ x_1 < x_2 }
\left[ f_{\phi\overline\phi}(x_1,x_2)
|x_1,x_2\rangle_{\phi\overline\phi} +
f_{\overline\phi\phi}(x_1,x_2)
|x_1,x_2\rangle_{\overline\phi\phi}\right]\;,\nonumber \\
\end{eqnarray}
where,
\begin{eqnarray}
f_{\overline Y}(x_1) &=& A_{\overline Y} e^{i (k_1 + k_2) x_1}\;, \\
f_{\phi\overline\phi}(x_1,x_2) &=& A_{\phi\overline\phi}(12)
e^{i(k_1 x_1 + k_2 x_2)} +
 A_{\phi\overline\phi}(21)
e^{i(k_2 x_1 + k_1 x_2)} \;, \\
 f_{\overline\phi\phi}(x_1,x_2) &=& A_{\overline\phi\phi}(12) e^{i(k_1 x_1 + k_2 x_2)} +
 A_{\overline\phi\phi}(21)
e^{i(k_2 x_1 + k_1 x_2)}\;.
\end{eqnarray}

By solving for the amplitudes we get,

\begin{eqnarray}
\label{S2}
\left(%
\begin{array}{c}
  A_{X\overline X}(12) \\
  A_{\overline X X}(12) \\
  A_{Z\overline Z}(12) \\
  A_{\overline Z Z}(12) \\
\end{array}%
\right) = \left(%
\begin{array}{cccc}
  \alpha & \beta & \gamma & \gamma \\
  \beta & \alpha & \gamma & \gamma \\
  \gamma & \gamma & \alpha & \beta \\
  \gamma &  \gamma & \beta & \alpha \\
\end{array}%
\right)\left(%
\begin{array}{c}
   A_{X\overline X}(21) \\
  A_{\overline X X}(21) \\
  A_{Z\overline Z}(21) \\
  A_{\overline Z Z}(21) \\
\end{array}%
\right)\;,
\end{eqnarray}
where
\begin{eqnarray}
\alpha(k_2,k_1) &=& - \frac{{\left( -1 + e^{i \,{k_1}} \right)
}^2\,{\left( -1 + e^{i \,{k_2}} \right) }^2}
    {\left( 1 + e^{i \,{k_2}}\,\left(e^{i \,{k_1}}  -2   \right)  \right) \,
      \left( 1 + e^{i \,{k_1}}\,\left( e^{i \,{k_2}} -2   \right)  \right) }\;, \\
\beta(k_2,k_1) &=& - \frac{{\left( e^{i \,{k_1}} - e^{i \,{k_2}}
\right) }^2}
    {\left( 1 + e^{i \,{k_2}}\,\left( e^{i \,{k_1}} -2  \right)  \right) \,
      \left( 1 + e^{i \,{k_1}}\,\left(e^{i \,{k_2}}  -2   \right)  \right) } \;,  \\
\gamma(k_2,k_1) &=& - \frac{\left( -1 + e^{i \,{k_1}} \right)
\,\left( e^{i \,{k_1}} - e^{i \,{k_2}} \right) \,
      \left( -1 + e^{i \,{k_2}} \right) }{\left( 1 + e^{i \,{k_2}}\,\left(e^{i \,{k_1}}  -2   \right)  \right) \,
      \left( 1 + e^{i \,{k_1}}\,\left( e^{i \,{k_2}} -2   \right)  \right) }\;.
\end{eqnarray}

One can put together both results above to obtain the $16\times 16$ matrix which corresponds to all the possible $2\to 2$ scattering processes.

A necessary and sufficient condition for the integrability of the
Hamiltonian we have calculated is the factorization of any
multiple particle scattering process in terms of two-body
scattering matrices, which leads to the Yang-Baxter equation for
three impurity systems \cite{intsys, qgroups}. The Yang-Baxter
equation can be easily verified with the above $S$-matrix.

 The necessary and sufficient condition for the integrability of
the boundary conditions of the open spin chain is the equality of
the S-matrix corresponding to the two processes shown in the
following figure \ref{fig:BYB.eps} \cite{intsys, qgroups}.

\myfig{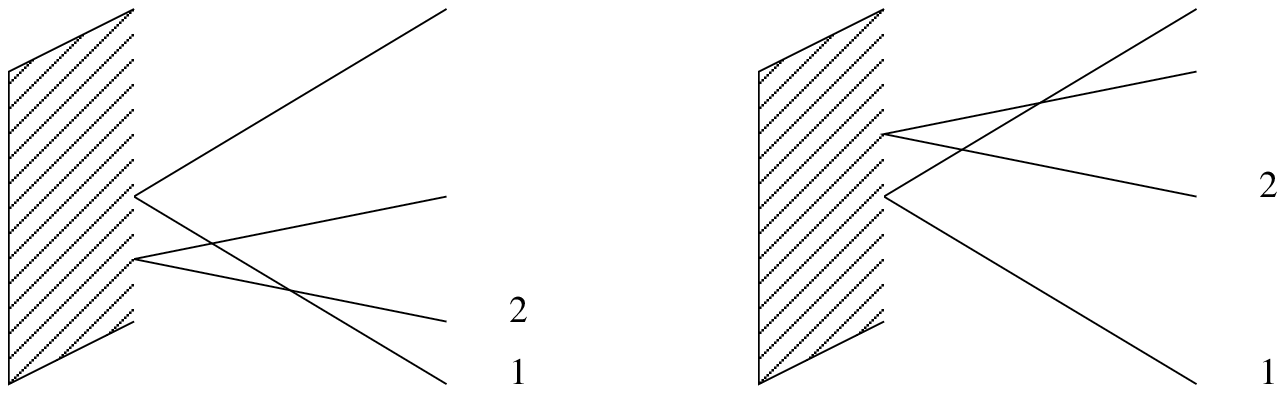}{9}{Pictorial representation of the Boundary Yang Baxter equation. The lines represent defects scattering from each other and the boundary (which is represented by the shaded region) in different orders.}

In our notation this translates to the following boundary Yang
Baxter equations:
\begin{eqnarray}
S^{\gamma_1 \gamma_2}_{ij}(-k_1,-k_2) K^{\gamma_3}_{\gamma_1}(k_2)
S^{\gamma_4 l}_{\gamma_3 \gamma_2}(k_2, -k_1)K^k_{\gamma_4}(k_1) =
K^{\gamma_1}_i(k_1)S^{\gamma_2 \gamma_3}_{\gamma_1
j}(k_1,-k_2)K^{\gamma_4}_{\gamma_2}(k_2)S^{k l}_{\gamma_4
\gamma_3}(k_2,k_1)\;,\nonumber \\
\end{eqnarray}
for the left boundary, and
\begin{eqnarray}
S^{\gamma_2 \gamma_1}_{ij}(-k_1,-k_2) K^{\gamma_3}_{\gamma_1}(k_1)
S^{k \gamma_4}_{\gamma_2 \gamma_3}(-k_2, k_1)K^l_{\gamma_4}(k_2) =
K^{\gamma_1}_j(k_2)S^{\gamma_3 \gamma_2}_{i
\gamma_1}(-k_1,k_2)K^{\gamma_4}_{\gamma_2}(k_1)S^{k l}_{\gamma_3
\gamma_4}(k_2,k_1)\;,\nonumber \\
\end{eqnarray}
for the right boundary. Using the bulk S-matrix derived in the
last section and using the boundary reflection matrix (\ref{bSm})
it is straightforward to verify that the boundary Yang Baxter
equations are indeed satisfied.

\section{Discussion}

We have found that maximal giant gravitons give rise to integrable boundary
conditions for the $SO(6)$ sector of the $PSU(2,2|4)$ open spin chain that corresponds to the dual operators of open strings attached to the maximal giant graviton.

There is an obvious generalization of this calculation which give rise to integrable boundary conditions. This is to consider the same calculation in the context of an orbifold of ${\cal N}=4 $ SYM in the presence of certain dibaryon operators (these type of dibaryons were considered orginally in  \cite{GRW}). It turns out that the calculation in abelian orbifold models in the presence of a single dibaryon gives the same answer we found. Dibaryon states can be more complicated in non-abelian orbifolds, this is why the analysis is harder to perform.
 As shown in various papers (see \cite{Orbs} and references therein), orbifold models correspond to the same spin chain model with possibly twisted boundary conditions (this accounts for discrete torsion).
 This twisting only manifests itself in the so-called hopping terms, where defects change position. In our case, at
 the boundary there is no hopping, because we don't exchange the field at the boundary
 with the $Z$ that make up the giant graviton. This means that the answer we would obtain in this case is identical to what we already have and does not need to be calculated again. This is exactly what one would expect from how D-branes are constructed in an orbifold in terms of image branes in the ${\cal N}= $ SYM theory. In this case all the image branes are maximal giant gravitons located on top of each other, and one expects that the boundary conditions in this case don't change at all.

It is more interesting to consider smaller giant gravitons. These require a more involved combinatorial treatment than what we have used here. Also, as discussed in the text, these giant gravitons are not static in global coordinates of $AdS_5\times S^5$. They are moving objects. From this point of view, a string attached to them is dragged by the giant graviton,
and it acquires some momentum in the same direction than the giant graviton is moving. in principle this momentum can be fractional .
This effect has to be taken into account before we can attempt to describe the same physics we have described here. To obtain an effective fractional momentum, the string must be exchanging momentum with the giant graviton. This means that the string can absorb and emit
partons at the boundary. This implies that the length of the spin chain is not fixed. This is
very important because a similar observation has been made for the $PSU(2,2|4)$ spin chain beyond one loop \cite{SU23}. This has been an obstacle to understanding the integrability beyond one loop and has led to very interesting proposals for an all loop result based on integrability and further assumptions \cite{Novel}. We are currently investigating this issue \cite{BCV}

\section*{Acknowledgments}

D. B. would like to thank S. Cherkis for many discussions on
integrable spin chains, and V. Balasubramanian, B. Feng and M.
Huang for many discussions on giant gravitons. D.B. work supported
in part by a DOE outstanding Junior Investigator award, under
grant DE-FG02-91ER40618  and National Science Foundation under
Grant No. PHY99-0794. S.V. work was supported by an NSF Graduate
Research Fellowship.

\end{document}